\documentstyle[psfig]{mn}

\begin{document}

\title [ The intermediate polar RW UMi]
{ Is the old nova RW UMi (1956) an intermediate polar? }

\author[A. Bianchini et al. ]
{Antonio Bianchini$^{1,2}$,
 Claus Tappert$^{2,3}$,
 Heather Osborne$^{1,4}$,
Ronald Canterna$^1$\cr and
Fabrizio Tamburini$^5$\\
$^1$ Dept. of Physics and Astronomy, University of 
Wyoming, USA\\ 
$^2$ Department of Astronomy, University of Padova, Italy \\
$^3$ Departamento de F\'{\i}sica, Universidad de Concepci\'on, Chile\\
$^4$ Astronomy Department, New Mexico State University, Las Cruces, USA \\
$^5$ Relativity and Cosmology Group, Portsmouth University, Hampshire Terrace, 
 PO 21EG, Portsmouth, UK }

\maketitle

\begin{abstract}
From photometric observations of the old nova RW UMi (1956) we have
discovered the presence of five sub-orbital periodicities. The longest period, 
$P_1 = 54.4$ min, always observed,  could be the beat period between the  
orbital period and the white dwarf (WD) spin period that we assume to be 
$P_2 = 33.4$ min. The other suborbital periods could be  orbital sidebands with respect to $P_2$ and/or correspond to Keplerian resonance radii.
We suggest that in its post-outburst state,  characterized by higher $\dot{M}$, 
RW UMi is an asynchronous intermediate polar with a truncated accretion disc.  
If the magnetic momentum of the WD is $\mu_{34} \sim 0.8$,
in its quiescent pre-outburst state,  the nova would appear as a discless, synchronous intermediate polar.

\end{abstract}


\section{Introduction}

Classical novae (CNe) belong to the class of Cataclysmic Variables (CVs) and
are formed by a white dwarf (WD) primary accreting matter from a Roche lobe 
filling companion, usually a lower main sequence star. The WD is surrounded by 
an accretion disk, unless its magnetic field is strong enough to partially 
({\it Intermediate Polar} systems) or totally ({\it Polar} systems) control the
accretion geometry. Stable mass transfer is ensured by angular momentum losses 
from the binary system due to the magnetic braking mechanism and gravitational 
radiation. Thus, CVs evolve towards shorter and shorter orbital periods and 
lower and lower masses of the secondary. 

The orbital periods of CVs show a bimodal distribution characterized by a 
minimum period of $P\sim80$ min and  a gap in the range 2\,h $\leq P_{\rm orb} 
\leq$ 3\,h (see Warner 1995 for a comprehensive review). The 3\,h 
upper limit of the gap  corresponds to  secondary stars which are just becoming
fully convective ($M_2 \approx 0.3\,M_\odot$) and in which the dynamo mechanism
is partially suppressed leading to a drastic decrease of the efficiency of the 
magnetic braking and, consequently, of the accretion rate \cite{zang+97}. The 
secondary then shrinks inside its Roche lobe until gravitational radiation will
bring them again in contact, which will happen when the binary has reached 
$P_{\rm orb}\sim$ 2\,h.  

The accretion luminosities of old novae are at any given orbital period 
systematically brighter than those of dwarf novae (DNe), suggesting that the
former ones have higher mass transfer rates and/or hotter white dwarfs and
thermally stable accretion discs. There seems to be no systematic difference 
between the pre- and the post- nova luminosities, although a few old novae
appear 2--4 magnitudes brighter than in their pre-nova state. Amongst these, 
GQ Mus, CP Pup and V1974 Cyg have orbital periods below the period gap, while
V1500 Cyg is just above it \cite{rittkolb98}. However, while strongly 
magnetized  systems like GQ Mus and V1500 Cyg, as  `polars', are expected to 
alternate between high and low states of their accretion rates, the explanation
for the luminosity standstills of disc accreting systems like CP Pup and V1974 
Cyg suggested by Retter \& Naylor \shortcite{rettnayl00} is that, after the 
outburst, they experience a constant superhump phase.

RW UMi is a high galactic latitude, probably slow, nova that brightened in 1956 
reaching $m\sim 6$. It was first considered as a supernova by Kukarkin 
\shortcite{kuka62} due to its unusually large outburst amplitude and because it 
was apparently located out of the Galactic halo. The pre-nova magnitude was 
estimated $\sim 21$ \cite{duer87} but its post-outburst magnitude still is 
about 2.2 mag brighter. Esenoglu et al.\ \shortcite{esen+00} estimated the 
nebular parallax of the nova as $5000 \pm 800$ pc. This yields 
$M_V^{\rm max}\sim-7.7$ at light maximum, and $M_V^{\rm min}\sim7.3$ or 
$\sim5.1$ before and after the outburst, respectively. Thus the nova is located
in the Galactic halo. 

The first spectroscopic observation of RW UMi was obtained by Zwicky 
\shortcite{zwic65} seven  years after the maximum. The spectrum shows a bluish 
continuum with emissions of hydrogen, HeI and HeII. Subsequent spectra obtained
by Kaluzny \& Chlebowski \shortcite{kaluchle89}, Szkody \& Howell 
\shortcite{szkohowe92} and Ringwald et al.\ \shortcite{ring+96} confirmed the 
steep blue continuum and the relatively high excitation state of the object.

The photometric behaviour of RW UMi at light minimum has been object of debate.
No light variations were found by Kaluzny \& Chlebowski \shortcite{kaluchle89};
Szkody et al.\ \shortcite{szko+89} suggested a periodicity of the order of 3 h,
while Howell et al.\ \shortcite{howe+91} found a period of 1.88 h. The 
possibility that the orbital period of RW UMi falls below the 2--3 h gap of CVs 
\cite{warn95} triggered more systematic observations until Retter \& Lipkin 
\shortcite{rettlipk01} were able to unambiguosly determine a photometric period
of $0.05912\pm0.00015$ d with a 0.025 mag amplitude. In the following we 
will refer to this period as the {\em main} period $P_{\rm main}$, as its 
precise nature is not really clear, although it is probably closely related to 
the orbital motion.

An orbital period of the order of 85 min definitely places RW UMi well below 
the 2 h lower limit of the period gap. This region of the Period-Luminosity 
diagram of CVs is populated by SU UMa systems. Retter \& Lipkin 
\shortcite{rettlipk01} suggested that RW UMi is a non-magnetic CV which is
experiencing a long post-outburst superhump phase like two more 
short-orbital-period old novae: CP Pup and V1974 Cyg.

In this paper we present the results of a photometric study of the old nova 
aimed to identify possible sub-orbital periodicities in its light curve. We 
have obtained CCD images in 1999 and in 2000 using the Wyoming Infrared 
Observatory (WIRO) and the  Ekar (Asiago) telescope, respectively. In section 2
we present the observation and the data reduction, and present the photometric 
periods discovered. In section 3 we discuss the general photometric behaviour 
of the old nova and suggest a physical explanation for the sub-orbital periods 
observed. The possible intermediate polar state of the post-nova is also 
discussed.

\section{The observations and data reduction}

\begin{figure}
\psfig{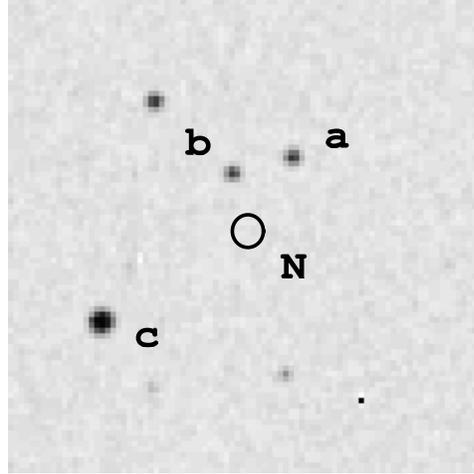}
\caption{ $2.5'' \times 2.5'' $ field of nova RW UMi. North is at the top,
East is to the left. The position of the nova is marked by a circle 
but the object is invisible ($m_v \ge 21$) in the POSS before its eruption 
in 1956.
Stars {\bf a} and {\bf b} have been used to obtain the mean relative fluxes 
of the nova observed with WIRO. }  
\label{field} 
\end{figure}

\begin{table}
\caption{\label{rwobs} The journal of the observations}
\begin{center}
\begin{tabular}{ccccc}
\hline
JD&      Telescope&   Filter& $ \Delta T$& $\delta T$  \\ 
start&   &                & (h)      &  (s)   \\
\hline
2451347.65930  &   WIRO& clear & 2.723&   23  \\
2451348.59151  &   ``   &  ``  & 2.604&   `` \\
2451695.45129 &    Ekar&    $V$&     3.227&   380   \\
2451696.46297  &     ``&    ``&  3.195      &  `` \\
\hline 
\end{tabular}
\end{center}
\end{table}

The $2.5'' \times 2.5'' $ field centered on the  nova is shown in Fig. 
\ref{field}. No object is seen at the position of the nova because the POSS 
image was taken before the 1956 outburst. However, the luminosity of the 
post-nova is higher than the pre-nova, and its image, together with the 
magnitudes and colours of two  field stars, labelled {\bf b} and  {\bf c} in 
Fig. \ref{field}, were provided by Kaluzny \& Chlebowski 
\shortcite{kaluchle89}. These field stars allowed an absolute calibration of 
the Ekar $V$ magnitudes. Stars {\bf a} and {\bf b} in Fig. \ref{field} were 
used to obtain the mean relative fluxes of the nova from the WIRO observations. 
The journal of observations is given in Table \ref{rwobs}.

\subsection{ WIRO observations}

\begin{figure}
\psfig{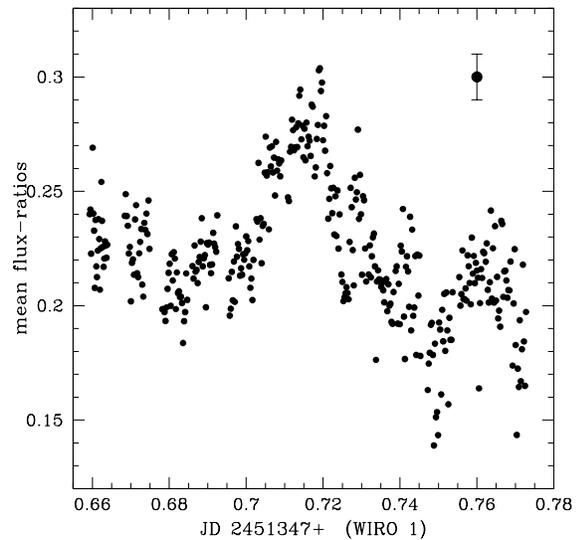}
\caption{Mean flux ratios of RW UMi with respect to field stars {\bf a} and 
{\bf b} of Fig.\ \ref{field} obtained during the first night at WIRO. No filter
was used. The typical error bar of individual data points is shown in the upper
right of the plot.}  
\label{wiro1} 
\end{figure}

\begin{figure}
\psfig{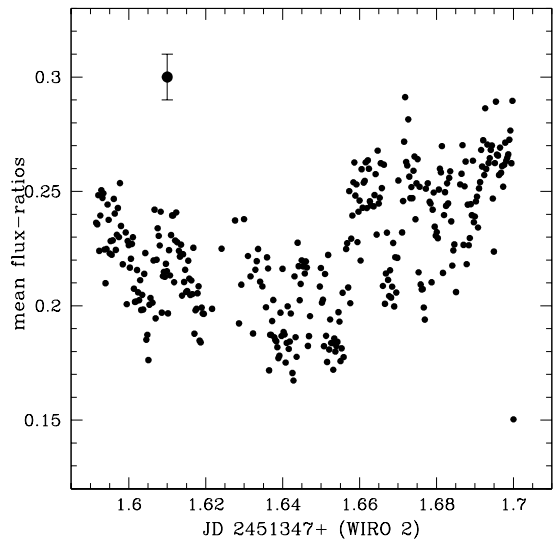}
\caption{Mean flux ratios observed on night 2 at WIRO; same as in Fig. 
\ref{wiro1}. The main modulation is not obvious in this data set. }  
\label{wiro2} 
\end{figure}

WIRO observations were secured in June 1999 using the Michigan State University visual camera attached at the prime focus of the 2.3 m telescope (WIRO) on Jelm
Mountain (Laramie, WY). The space resolution was 2.3 arcsec/pixel. We observed
with a time resolution of 27 s for 2.72 h on June 18, and for 2.60 h on June 
19. In order to obtain a higher S/N ratio we did not use filters. We selected a
small region for the CCD readout in order to improve the time resolution of our
observations. The selected field therefore contained only RW UMi and field 
stars {\bf a} and {\bf b} (see Fig. \ref{field}). We then measured the mean 
flux ratios between the nova and the two nearby stars following standard 
photometric procedure. The light curves of the nova, expressed as flux ratio 
variations,  for the two observing nights at WIRO are shown in Figs. 
\ref{wiro1} and \ref{wiro2}. We note that, in spite of the fact that each run 
was about 1.9 times the main period, its modulation is not clearly seen in our 
data. The reason is that the light curve of the nova is intrinsically rather 
irregular which also explains why only the systematic and extended observations
performed by Retter \& Lipkin \shortcite{rettlipk01} were able to provide a 
precise period.

\subsection{Ekar observations}

\begin{figure}
\psfig{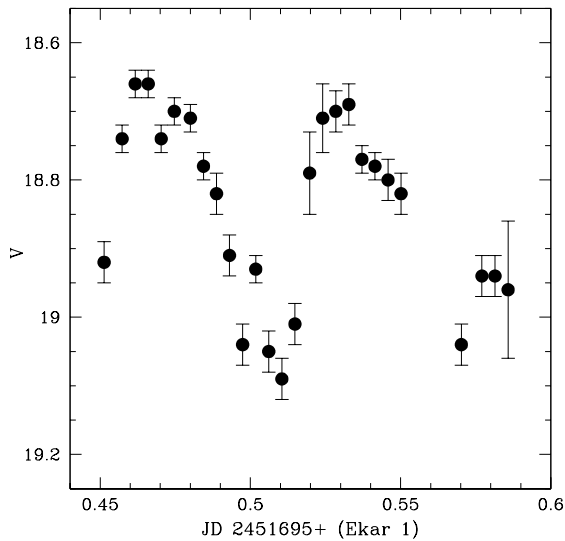}
\caption{$V$ magnitudes and error bars of RW UMi obtained during the first 
observing night at Mt.\ Ekar.}  
\label{ekar1} 
\end{figure}

\begin{figure}
\psfig{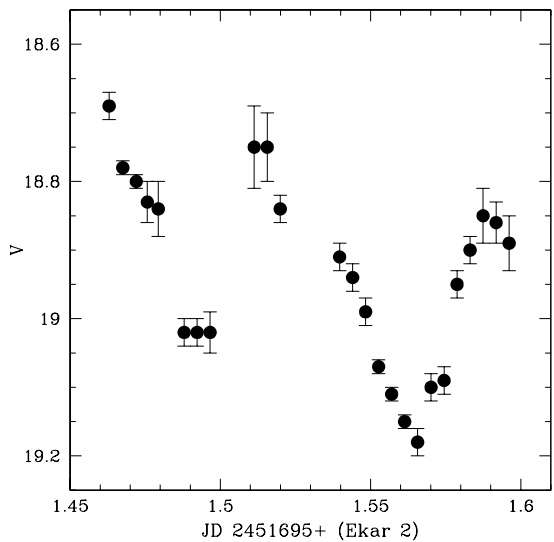}
\caption{Mt.\ Ekar $V$ magnitudes obtained on night 2.   }  
\label{ekar2} 
\end{figure}

Additional observations in $V$ were obtained in May 2000 at Asiago, Italy, 
using the 1.82 m Mt. Ekar telescope with the CCD of the AFOSC system as an 
imager. The time resolution was 430 s. Data reduction was performed with IRAF 
software packages. Field stars {\bf b} and  {\bf c} of Fig. \ref{field} allowed
the calibration of the $V$ magnitudes \cite{kaluchle89}. The light curves of 
the old nova from the two Ekar observing nights are shown in Figs.\ \ref{ekar1}
and \ref{ekar2}. The length of each observation is about 2.3 orbital periods,
and in this case the orbital modulation is more evident than in the WIRO data.

\subsection{The photometric periods.}

\begin{table}
\caption{\label{rwpeaks} Suborbital periodicities found in RW UMi. The specific
values were determined by subtracting all other frequencies but the one under
consideration. They are here presented with a precision which only reflects the
resolution of the periodogram. The systematic uncertainty due to the broadness
of the individual peaks is much larger and estimated to amount to $\sim \pm2$
d$^{-1}$. The main period $P_{\rm main}$ determined by Retter \& Lipkin 
(2001) is also reported: according to them it corresponds to a positive 
superhump period.}
\begin{center}
\begin{tabular}{ccccc}
\hline
 Telescope & Night & Peak & Frequency  & Period   \\ 
           &       &      & (d$^{-1}$)   & (min)   \\
\hline
 WIRO       & 1+2  & $P_1$    & 26.447     & 54.449   \\
            &  1    & $P_2$   & 43.150     & 33.372   \\
            &  2    & $P_3$   & 59.000     & 24.407   \\
            & 1+2  & $P_4$    & 82.820     & 17.387   \\
\hline
 Ekar       & 1+2  & $P_1$    & 26.50      & 54.34    \\
            &  1   & $P_2$   & 43.00      & 33.49    \\
            &  1   & $P_5$   & 68.10      & 21.14    \\
\hline
main     &    & $P_{\rm main}$ & 16.915     & 85.133   \\
period      &    &           &            &          \\
\hline
\end{tabular}
\end{center}
\end{table}

\begin{figure}
\psfig{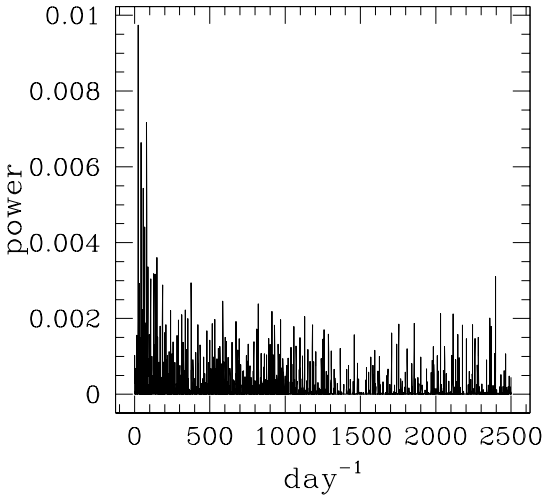}
\caption{{\sc clean} power spectrum of all WIRO data up to their Nyquist 
frequency, after subtraction of the orbital modulation and long term trends: 
significant signals are below frequency 200 d$^{-1}$. }  
\label{clean12} 
\end{figure}

\begin{figure}
\psfig{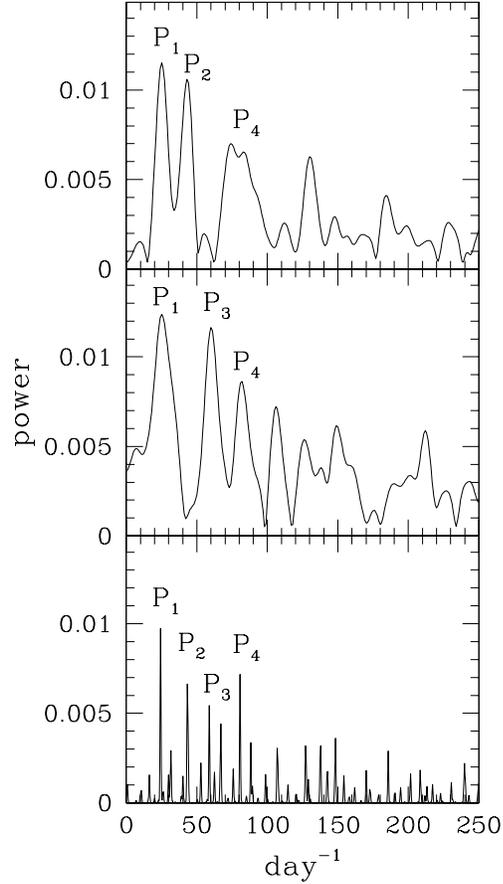} 
\caption{{\sc clean} power spectra of WIRO data after correction for the 
long-term trends. {\it Upper panel}: night 1 shows three main peaks, $P_1$, 
$P_2$ and $P_4$ at about 23, 43 and 80 d$^{-1}$, respectively; {\it medium 
panel}: night 2 still shows  $P_1$ and $P_4$  but also another consistent 
signal, $P_3$, at about 60 d$^{-1}$; {\it lower panel}: in the combined data of
nights 1 and 2, $P_1$ and $P_4$ appear more prominent than  $P_2$ and $P_3$ 
because they were observed in both nights. }
\label{clean3} 
\end{figure}

\begin{figure}
\psfig{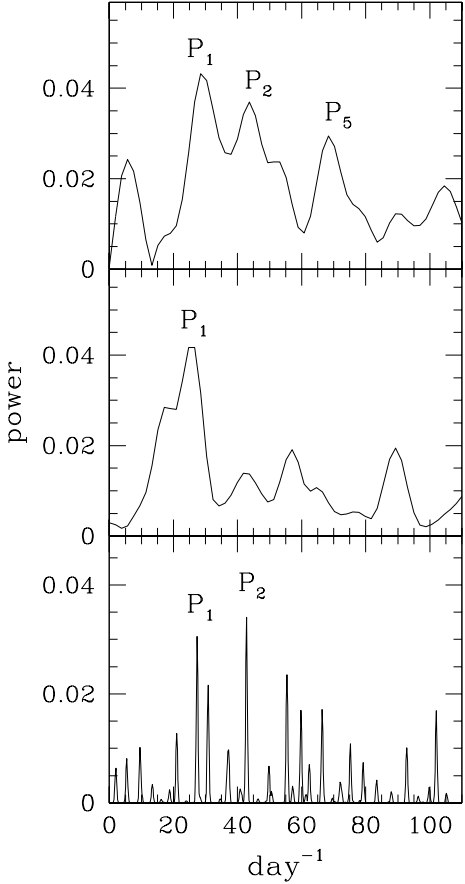} 
\caption{{\sc clean} power spectra of Ekar observations after correction for 
the long-term modulation. {\it Upper panel}: night 1 shows peaks around 28 
d$^{-1}$, relatively close to $P_1$, at 43 d$^{-1}$, which is $\sim P_2$, and 
around 67 d$^{-1}$ ($P_5$); {\it medium panel}: on night 2 only a broad peak 
around 25 d$^{-1}$, that is $P_1$, is relevant; {\it lower panel}: in the 
combined data of nights 1 and 2,  $P_1$ and $P_2$ are still prominent. }
\label{clek3} 
\end{figure}

\begin{figure}
\psfig{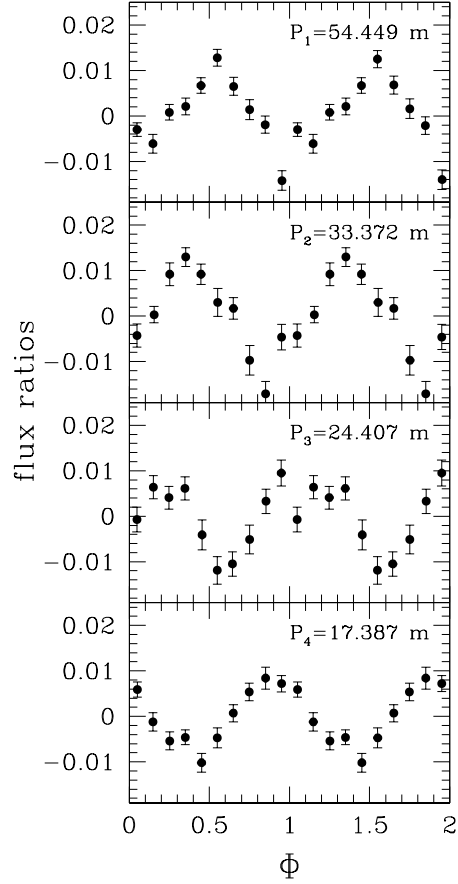} 
\caption{WIRO data folded with the four main signals detected. For the folding
we used the peak values of the periodogram indicated in each panel. Note that 
the given precision only corresponds to the resolution of the periodogram 
(see also Table \ref{rwpeaks}). Individual points and error bars refer to the
means within each $0.1P$ phase interval. Phase zero refers to the first 
datapoint (see Table \ref{rwobs}). $P_1$ and $P_2$ are the more stable signals 
(see also their smaller errorbars). }
\label{wplot4} 
\end{figure}

\begin{figure}
\psfig{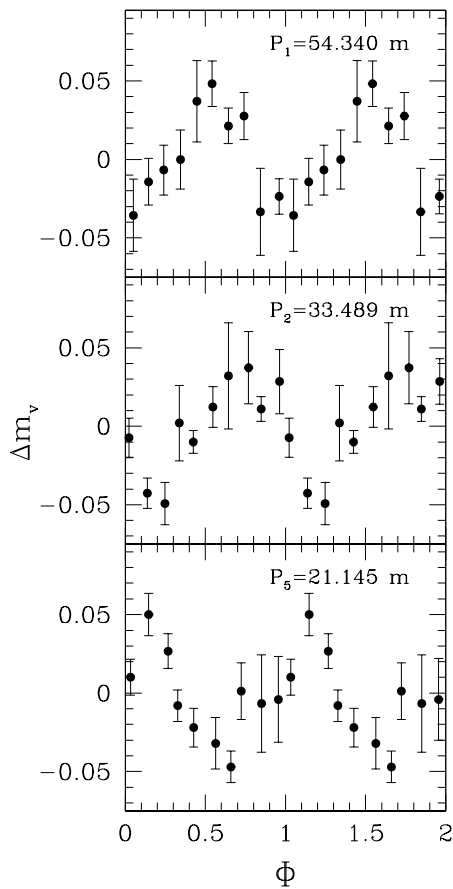} 
\caption{ Ekar data folded  with the three main periods detected and reported 
in Table \ref{rwpeaks}. $P_1$ is observed in both nights; $P_2$ is detected
only in night 1 as well as  period $P_5$. See also comments to 
Fig.\ \ref{wplot4}.}
\label{eplot3} 
\end{figure}

As anticipated, the pathological irregularity of the light curve of this old 
nova makes the determination of the main period \cite{rettlipk01} difficult, 
when only two or three cycles are observed. For example, the Fourier anlysis of
the WIRO data, probably due to the rather different appearance of the two data 
sets, would suggest a  period of 0.098 d. The Ekar data, instead, show a best 
period at 0.06614 d, while a peak at 0.05856 d, the closest one to 
$P_{\rm main}$, is only the fifth highest alias in the power spectrum. 

In order to look for the presence of higher frequency secondary modulations, we
have corrected our data sets subtracting the observed orbital modulation and 
long term trends. We performed harmonic analysis using the {\sc clean} program 
which eliminates aliases \cite{robe+87}. Power spectra are presented in Figs. 
\ref{clean12}--\ref{clek3}. Fig.\ \ref{clean12} shows that all the significant 
peaks in WIRO data are at frequencies below 200 d$^{-1}$. Fig. \ref{clean3} 
shows the results for individual nights. Night 1 shows three main periods, 
$P_1\sim54$ min,  $P_2\sim33$ min, and $P_4\sim17$ min; night 2 still shows 
$P_1$ and $P_4$, but also $P_3\sim24$ min.   

Fourier analysis of the two Ekar runs is shown in Fig.\ \ref{clek3}. A year 
after WIRO observations, $P_1$ is still the more stable period, while $P_2$ 
and a modulation $P_5\sim 21$ min are observed only in the first night.   

Each of the detected periodicities is represented by a folded light curve in
Figs. \ref{wplot4} and \ref{eplot3}. The periods adopted are those of 
Table \ref{rwpeaks} and do not coincide exactly with those of 
Figs.\ \ref{clean3} and \ref{clek3}. They were found by performing Fourier 
analysis on our data sets after the subtraction of all the other periods except
the one under examination. In this way, each modulation is no longer distorted 
by the simultaneous presence of the other periods. The small error bars suggest
that the modulations are rather well defined.

\section{Discussion}

\subsection{ The long term light curve}   

\begin{table*}
\caption{\label{rwmag} Historical magnitude determinations for RW UMi.}
\begin{center}
\begin{tabular}{lccccccl}
\hline
JD&    $V$& $\pm \Delta V$&  $B$& $\pm \Delta B$& $B-V$& $V-R$& Ref. \\ 
\hline
2246778.05 &  18.70 &     &     &    &        &      &
Kaluzny \& Chlebowski \shortcite{kaluchle89}\\
2447741.5 &  18.52 &     &     &    &   $-$0.06&   0.21&
Szkody \shortcite{szko94}\\
2447451.5 &   18.5 & 0.2&     &    &        &       &
Szkody et al.\ \shortcite{szko+89}\\
2447452.5&   18.8&   0.2&     &    &        &       &
Szkody et al.\ \shortcite{szko+89}\\
2447769.7&   18.9:&  0.13:& 18.9& 0.13&      &       &
Howell et al.\ \shortcite{howe+91}\\
2448117.5&   18.90&       &     &     & 0.10&     0.20&
Downes \& Duerbeck \shortcite{downduer00}\\
2450962.0&    18.95&       &     &     &     &         &
Ringwald et al.\ \shortcite{ring+96}\\
2451347.72&   18.9:&  0.23:&     &     &      &       &
WIRO data (white light)\\
2451348.65&   18.9:&  0.23:&     &     &      &       &
WIRO data (white light)\\
2451695.52&   18.93&   0.27&        &     &      &       &
Ekar data \\
2451696.52&   18.94&   0.24&        &     &      &       &
Ekar data\\
\hline 
\end{tabular}
\end{center}

\end{table*}

Table \ref{rwmag} collects some historical magnitude determinations of RW UMi,
and reports, whenever possible, the  range of the light modulation observed and
the colours. While this list suggests a slow negative trend of the luminosity 
of the old nova, the reported (mean) magnitudes all lie within the observed
short-term variations of the individual runs. Thus, only future observations 
will be able to confirm such a possible decline.

\subsection{Physical parameters of the binary}

Assuming that the period derived by Retter \& Lipkin \shortcite{rettlipk01} is 
very close to the actual orbital period (see next section), and taking the 
recent correlation between the orbital period and the mass of the secondary 
star presented by Howell \shortcite{howe01}, we may estimate the mass of the 
secondary in RW UMi as $M_2 \sim 0.115 M_\odot$.

In order to constrain the mass of the primary we make use of the velocity of
decline from the nova outburst. Warner \shortcite{warn95} reports $t_3$ = 
114 d, which corresponds to a decline velocity $v_{\rm dec}$ = 0.02 mag/d. 
Using the calibration given by della Valle \shortcite{dell91} we obtain a 
maximum absolute magnitude $M_{\rm max} \sim -7$ mag. According to Livio
\shortcite{livi92} this is the luminosity produced by a white dwarf primary 
with $M_1 \sim 0.7 M_\odot$. With this, we obtain a mass ratio 
$q = M_2/M_1 \sim 0.16$. 

The above calculated masses and main geometrical parameters of the binary
system are collected in Table \ref{system}.

\subsection{Superhumps and the orbital period}

CVs can show several photometric periods, the most important ones are related 
to the orbital period. The most obvious case is that of CVs at sufficiently
high inclination showing eclipses and/or luminosity bumps due to the hot spot.
Other important photometric modulations, which do not seem to be strongly 
related with the inclination of the binary system, are called `superhumps' and 
may differ by a few percent from the orbital period. Actually, two opposite 
cases are observed:

\smallskip \noindent
{\bf a)} {\it Positive superhumps}. They mostly characterize SU UMa DN systems 
during super outbursts and are thought to be produced at relatively large mass 
transfer rates, when the outer edge of the accretion disc expands reaching the 
3:1 resonance radius. Positive superhumps then represent the  beat period 
between the apsidal prograde precession of the eccentric outer edge of the disc
and the orbital motion of the secondary. 

Taking equations (3.34) and (3.42) from Warner \shortcite{warn95} we find a
relation between the positve superhump period $P_{\rm SH_+}$ and the orbital
one $P_{\rm orb}$:
\begin{equation}
 \frac{P_{\rm SH_+}}{P_{\rm orb}} = \frac{1+q}{1+0.74~q}~.
\end{equation}
With $q$ = 0.16, we obtain a positive superhump period 4\% larger than the
orbital one.

This situation corresponds to the scenario suggested by Retter \& Lipkin 
\shortcite{rettlipk01}. They also suggested that, similarly to the other two 
short-orbital-period old novae CP Pup and V1974 Cyg, the presence of positive 
superhumps might be a consequence of the post-outburst higher accretion
luminosity.

\smallskip \noindent
{\bf b}) {\it Negative superhumps}. A few CVs are known to show cyclical 
brightness modulations with periods shorter than the orbital one. This 
phenomenon is generally interpreted as the result of a precessing disc.
The precession of the outer anulus of a tilted disc in this case is retrograde. 
From Warner's \shortcite{warn95} equation (2.62c) we obtain for the negative
superhump period $P_{\rm SH_-}$:
\begin{equation}
 \frac{P_{\rm SH_-}}{P_{\rm orb}} = \left(1+0.35~\frac{q}{(1+q)^2}\right)^{-1}~.
\end{equation}
For RW UMi this results in $P_{\rm SH_-}$ being 4\% shorter than $P_{\rm orb}$.

\smallskip\noindent
We conclude that the period observed by Retter \& Lipkin \shortcite{rettlipk01}
should represent the orbital period within 4\% uncertainty. We furthermore note
that our derived mass ratio $q$ = 0.16 well satisfies the condition $q \le$ 
0.22 for the formation of positive superhumps. This might favour the 
interpretation by Retter \& Lipkin \shortcite{rettlipk01} that $P_{\rm main}$
represents a positive superhump period.

\subsection{ The suborbital photometric periods}

\begin{table}
\caption{\label{system} Dynamical and geometrical parameters of the binary
system. }
\begin{center}
\begin{tabular}{cccc}
\hline

$P_{\rm main}^{(1)}$ & $M_1^{(2)}$  & $q^{(3)}$     & $a^{(4)}$ \\ 
 (d)                & ($M_{\odot}$)& $M_{2}/M_{1}$ & ($10^{10}$ cm)   \\
0.05912             & 0.7          &  0.16         & 4.16 \\
\hline
 $R_{\rm L1}^{(5)}$ & $R_{\rm WD}^{(6)}$ &    & \\ 
  ($10^{10}$ cm)    & ($10^8$ cm)        &    &   \\
2.83                & 5.655              &    & \\
\hline
\multicolumn{4}{l}{ 1) is in all cases close to $P_{\rm orb}$ (see text);}\\
\multicolumn{4}{l}{ 2) tentatively assumed  for a slow nova; }\\
\multicolumn{4}{l}{ 3) $M_2$ is derived from Howell \shortcite{howe01}; }\\
\multicolumn{4}{l}{ 4) $a$ is the orbital separation; }\\
\multicolumn{4}{l}{ 5) Roche lobe radius of the WD \cite{warn95};}\\
\multicolumn{4}{l}{ 6) from Hamada \& Salpeter \shortcite{hamasalp61}. }\\
\end{tabular}
\end{center}
\end{table}

\begin{table}
\caption{\label{radper} Radii corresponding to the observed secondary periods.}
\begin{center}
\begin{tabular}{cccc}
\hline
 Period    &Symbol& Radius                & Notes   \\ 
  (min)    &      & ($10^{10}$ cm)        &         \\
\hline
$P_1$      &$r_1$ & 2.924  &    1    \\
$P_2$      &$r_2$ & 2.113  &     2    \\
$P_3$      &$r_3$ & 1.714  &     3    \\
$P_5$      &$r_5$ & 1.557  &     4    \\
$P_4$      &$r_4$ & 1.367  &     5    \\
\hline
\multicolumn{4}{l}{ 1) larger than $R_{L1}$; possible beat period} \\ 
\multicolumn{4}{l}{ between $P_2$ and $P_{\rm main}$; }\\
\multicolumn{4}{l}{  2) close to $r_{\rm max}$; possible spin period of the WD;}
\\
\multicolumn{4}{l}{  3) close to the superhump  resonance 3:1,}\\
\multicolumn{4}{l}{ ~~~~ i.e. the (3,2) resonance radius; }\\
\multicolumn{4}{l}{  4) close to the (4,3) resonance radius; }\\
\multicolumn{4}{l}{  5) $\sim 0.5 \times P_2$ }\\
\end{tabular} 
\end{center}
\end{table}

\begin{table}
\caption{\label{disc} Critical radii in the accretion disc of RW UMi.}
\begin{center}
\begin{tabular}{ccc}
\hline
 Symbol     & Radius                & Notes   \\ 
            & $10^{10}$ (cm)        &         \\
\hline

$r_{\rm circ}$  & 0.780  &     1    \\
$r_{43}$    & 1.571  &     2    \\
$r_{32}$    & 1.903  &     3    \\
$r_{21}$    & 2.494  &     4    \\
$r_{\rm max}$   & 2.152    &     5,6    \\
\hline
\multicolumn{3}{l}{  1) circularization radius \cite{hesshopp90};}\\
\multicolumn{3}{l}{  2) (4,3) resonance radius;     }\\
\multicolumn{3}{l}{  3) (3,2) resonance  radius responsible for superhumps;}\\
\multicolumn{3}{l}{  4) (2,1) outer resonance radius; it is larger than $r_{\rm max}$;}\\
\multicolumn{3}{l}{ ~~~~ sometimes responsible for spiral structures; }\\

\multicolumn{3}{l}{  5) last outer stable orbit \cite{pacz77}; }\\
\multicolumn{3}{l}{  6) positive superhumps require that the }\\
\multicolumn{3}{l}{ ~~~~ outer disc radius  reaches  $r_{32}$. }\\
\end{tabular}
\end{center}
\end{table}

Besides the superhumps, which periods are close to the orbital one, CVs may 
show other periodic or quasi-period oscillations (QPOs) in their light curves.
These phenomena might be similar to those observed in LMXBs. However, probably
due to the longer timescales in CVs, they have not been studied as thoroughly. 
The secondary, suborbital, periodicities found in RW UMi are listed in Table 
\ref{rwpeaks}.  

Lai \shortcite{lai99} has shown that the inner region of the accretion disc 
around a rotating magnetized collapsed object (neutron star, white dwarf) is 
subject to magnetic torques that induce warping and precession of the disc. The
result is that a number of magnetically driven resonances between the disc and 
the rotating central dipole will arise. In some intermediate polar systems 
(IPs) the spin period of the WD is the same as that of the Keplerian orbit at 
the magnetosphere radius, so the two rotational motions are equal. In this 
case, the magnetic field will easily produce a warped inner disc which rotates 
with the spin period of the WD. However, in order to produce observable 
interferences with the orbital motion, the warped structure must extend to a
large fraction of the disc, so that its variable geometry can interact with
the accretion stream from the secondary. This possibility will depend on the 
viscosity inside the disc and the intensity of the magnetic field of the 
primary. In general, outside some critical radius, viscous forces will overcome
the magnetic torque and warping will be destroyed \cite{lai99}.

However, since RW UMi is a short orbital period system, which can host only a 
very small accretion disc, even a modest magnetic field might produce a warping
that extends to a considerable portion of the disc. Still, most IPs are not 
synchronous rotators, so that the warping will be a much more
complicated process, in which the outer parts of the disc might switch between
prograde and retrograde precession. Thus, RW UMi might display both positive 
and negative superhumps, perhaps in a rather complex and unstable fashion. 

A key problem in the analysis of the suborbital periods is to detect the
spin period of the WD. From our data we note that the period $P_1$ 
(Table \ref{rwpeaks}) is the most stable one over at least one year. In terms 
of Keplerian velocities, it would correspond to an orbit with a radius 
$r_1 > R_{L1}$, which is obviously impossible. We however note that $P_1$ fits 
quite well with the beat period between $P_2$ and $P_{\rm main}$. Rather
curiously, this match is better than the one obtained using the two
possible alternative values for the orbital period $P_{\rm orb} = P_{\rm main}
\pm 4\%$ (see Section 3.3), although our precision does not really allow to 
conclusively decide between those possibilities.

$P_2$ was observed only during one night at WIRO and during one at Ekar. 
Although it is not the shortest of the detected periods, for the above reasons
we tentatively assume it as the spin period of the magnetic WD. In the most
general case of an inclined rotator, the inner disc region truncated by the 
magnetosphere of the primary should be warped and precess with the WD spin 
period. However, the most easily observable effect is the direct interaction of
the mass stream from the secondary with the rotating magnetosphere of the WD. 
The changing inclination of the magnetic axis with respect to the stream 
produces a periodic modulation of the accretion rate onto the polar caps. In 
this way $P_1$ results as the beat period between the orbital and the 
rotational one ($P_2$).

Following this scenario, we should also look for other optical orbital 
sidebands like $\omega_{\rm rot}+\Omega_{\rm orb}$ and $\omega_{\rm rot} \pm
2~\Omega_{\rm orb}$, where $\omega_{\rm rot} = P_2^{-1}$ and $\Omega_{\rm orb}
= P_{\rm main}^{-1}$. We find that
\begin{itemize}
\item $\omega_{\rm rot}+\Omega_{\rm orb} = 60 \sim P_3^{-1}$;
\item $\omega_{\rm rot}+2~\Omega_{\rm orb} = 77$, between $P_4^{-1}$ and
$P_5^{-1}$;
\item $\omega_{\rm rot}-2~\Omega_{\rm orb} = 9$.
\end{itemize}
The latter, low frequency value corresponds to the time length of our 
observations, and is therefore not detectable. However, we note that a peak of 
minor significance at roughly 8.6 d$^{-1}$ is present in the
power spectrum published by Retter \& Lipkin \shortcite{rettlipk01}.

Taking a different approach, we might try to interprete the observed secondary 
periods in terms of inhomogeneities in the disc. We therefore compute 
their Keplerian radii (Table \ref{radper}) and look for correlations with those
radii where matter streams are in resonance with the tidal action by the 
secondary. Table \ref{disc} lists the main critical radii of the accretion disc
for dynamical stability of orbiting matter streams. Radii labeled as $r_{ij}$ 
refer to resonances between the radial excursion of a test particle and the 
position of the secondary star \cite{WhiteKing91}. 

Periods $P_3$ and $P_5$ would indicate Keplerian radii $r_3$ and $r_5$ which 
are close to resonance radii $r_{32}$, where superhumps are thought to be
produced, and $r_{43}$, respectively. Finally, we note that $P_4$ is about 
$0.5\times P_2$ and could be its second harmonic.

If we assume that, in its steady quiescent state, the nova is
a synchronous rotator and take $P_2$ as the spin period of the
WD, then $r_2$ should coincide with the Alfven radius. However, this would suggest 
Keplerian orbits beyond the last stable orbit of the disc $r_{\rm max}$, and
no disc could form at all. As a result, if we assume that some disc does 
presently exist, then  RW UMi, in its post-outburst state, cannot be a syncronous rotator.

In order to have a standard disc, the stream from the secondary should be able 
to penetrate the magnetosphere of the primary down to the circularization 
radius $r_{\rm circ}$. Our data do not give us any clear indication about
what the inner radius of the disc can be. However, we might reasonably assume
that some material is presently orbiting around the WD at least between radii $r_3$ and $r_{32}$ which
accounts for the probable presence of superhumps.

\subsection{Pre-nova and post-nova state of RW UMi}

In the previous subsection we have tentatively assumed $P_2$ as the spin period
of the WD. If its Keplerian radius, $r_2$, is also the Alfven radius for the 
stream, $r_{\rm \mu,stream}$, then, as $r_{\rm \mu,stream} > r_{\rm circ}$, no 
disc can form. 

However, it appears plausible that a sort of an equilibrium status, with no 
disc formation, might have been achieved by the fainter pre-nova, while the 
stream was carrying only $\dot{M}\sim 10^{-9} {M}_\odot {\rm yr}^{1}$. Using 
equations (6.5) and (7.6) from Warner \shortcite{warn95} we may write 
\begin{equation}
r_{\rm \mu,stream}=3.66\times10^{10}\mu_{34}^{4/7}M_1^{-1/7}\dot{M}_{16}^{-2/7}
~{\rm cm}.
\label{mu_eq}
\end{equation}
Thus, within our hypothesis we derive $\mu_{34}\sim0.84$, or, consequently, 
$B=1.5\times10^7$ G.

The post-nova is however brighter than the pre-nova. Again we use equation 
(\ref{mu_eq}) to derive $r_{\mu,stream}$ for the high state of the nova, 
characterized by $\dot{M}\sim 10^{-8} {M}_\odot {\rm yr}^{-1}$, which would 
account for yielding $M_{\rm min}\sim 5.1$ mag. We find that the stream can now
penetrate the magnetosphere down to a radius 
\begin{displaymath}
r_{\rm \mu,stream}\sim 1\times10^{10} ~ {\rm cm} 
\end{displaymath}
This radius is probably not close enough to the circularization radius to allow
for the formation of a standard disc, but material can finally penetrate the 
magnetosphere ensuring stable accretion and also the formation of a probably 
chaotic disc. Note that the resonance radii of Table \ref{disc} are now 
populated by streaming gas. All intermediate polars have 
$r_{\rm inner}<r_{\rm corotation}$, so they are asynchronous rotators. The gas 
between the inner disc radius and the corotation radius might be rather 
turbulent and instability phenomena at the resonance radii could be enhanced. 

In conclusion, if we assume $P_2$ as the spin period of the WD, we find 
that  RW UMi in its steady quiescent state should be a discless intermediate polar, while in its post-outburst state it appears as an intermediate polar with
a somewhat unstable truncated accretion disc. The magnetic moment of the primary
$\mu_{34}\sim 0.84$ is rather large for short orbital period intermediate 
polars. RW UMi should then tend with time to go back to a pure discless state.

\section{Summary}

\begin{enumerate}
\item[1.] We have presented new photometric time-series of the old nova
RW Umi.
\item[2.] The data revealed the presence of several suborbital periods. The
most stable one of these is $P_1 \sim 54$ min, which is detected in all data
sets. We find that $P_1$ agrees well with a beat period between our period
$P_2 \sim 43$ min and the period found by Retter \& Lipkin 
\shortcite{rettlipk01} $P_{\rm main}$ = 85 min, which is assumed to be very
close to the orbital period. 
\item[3.] The most probable explanation for the presence of $P_2$ is that it
represents the spin period of the white dwarf. The other suborbital periods can
then be found as being optical orbital sidebands with respect to $P_2$ and/or
correspond to Keplerian resonance radii. 
\item[4.] Assuming a mixture of `disc' and `magnetic' periods we find that
RW UMi, in its post-nova state, must be an asynchronous intermediate polar,
as the Keplerian radius of the assumed spin period $P_2$, i.e. the Alfven 
radius, is larger than the circularization radius.
\item[5.] Comparing the mass-transfer rates from the pre- and the post-nova,
we conclude that RW UMi in its quiescent, i.e.\ pre-nova, state is a discless
intermediate polar.
\end{enumerate}

\section*{ACKNOWLEDGMENTS}
One of us (F.T.) wishes to thank the Department of Astronomy of the
University of Padova for the kind hospitality. 
This work has been supported by NSF REU  Grant AST 9732039, 
and by the Italian MURST.

\end{document}